\documentclass[fleqn,usenatbib]{mnras}

\usepackage{newtxtext,newtxmath}
\usepackage[T1]{fontenc}
\usepackage{float}

\DeclareRobustCommand{\VAN}[3]{#2}
\let\VANthebibliography\thebibliography
\def\thebibliography{\DeclareRobustCommand{\VAN}[3]{##3}\VANthebibliography}

\usepackage{graphicx}	% Including figure files
\usepackage{amsmath}	% Advanced maths commands
\usepackage{multirow}
\usepackage[dvipsnames]{xcolor}
\usepackage{orcidlink}

\title[NIRCam high redshift mergers]{The rate and contribution of mergers to mass assembly from NIRCam observations of galaxy candidates up to 13.3 billion years ago}

\author[Nicolò Dalmasso et al.]{Nicolò Dalmasso$^{1,2}$\thanks{e-mail: ndalmasso@student.unimelb.edu.au}\orcidlink{0000-0002-1850-4050},
Antonello Calabrò$^{3}$,
Nicha Leethochawalit$^{4}$
\orcidlink{0000-0003-4570-3159},
Benedetta Vulcani$^{5}$
\orcidlink{0000-0003-0980-1499},
Kristan Boyett$^{1,2}$,
\newauthor Michele Trenti$^{1,2}$
\orcidlink{0000-0001-9391-305},
Tommaso Treu$^{6}$
\orcidlink{0000-0002-8460-0390},
Marco Castellano$^{3}$,
Maru\v{s}a Brada\v{c}$^{7,8}$\orcidlink{0000-0001-5984-0395},
Benjamin Metha$^{1,2}$,
\newauthor Paola Santini$^{3}
\orcidlink{0000-0002-9334-8705}$
\\
$^{1}$School of Physics, University of Melbourne, Parkville, Vic 3010, Australia\\
$^{2}$Australian Research Council Centre of Excellence for All-Sky Astrophysics in 3-Dimensions, Australia \\
$^{3}$INAF Osservatorio Astronomico di Roma, Via Frascati 33, 00078 Monteporzio Catone, Rome, Italy\\
$^{4}$National Astronomical Research Institute of Thailand (NARIT), Mae Rim, Chiang Mai, 50180, Thailand\\
$^{5}$INAF Osservatorio Astronomico di Padova, vicolo dell'Osservatorio 5, 35122 Padova, Italy\\
$^{6}$Department of Physics and Astronomy, University of California, Los Angeles, 430 Portola Plaza, Los Angeles, CA 90095, USA\\
$^{7}$University of Ljubljana, Department of Mathematics and Physics, Jadranska ulica 19, SI-1000 Ljubljana, Slovenia\\
$^{8}$Department of Physics and Astronomy, University of California, Davis, 1 Shields Ave, Davis, CA 95616, USA
}

\date{Accepted XXX. Received YYY; in original form ZZZ}

\pubyear{2024}

\begin{document}
\label{firstpage}
\pagerange{\pageref{firstpage}--\pageref{lastpage}}
\maketitle

\begin{abstract}
We present an analysis of the galaxy merger rate in the redshift range $4.0<z<9.0$ (i.e. about 1.5 to 0.5 Gyr after the Big Bang) based on visually identified galaxy mergers from morphological parameter analysis. Our dataset is based on high-resolution NIRCam JWST data (a combination of F150W and F200W broad-band filters) in the low-to-moderate magnification ($\mu<2$) regions of the Abell 2744 cluster field.
From a parent set of 675 galaxies $(M_{U}\in[-26.6,-17.9])$, we identify 64 merger candidates from the Gini, $M_{20}$ and Asymmetry morphological parameters, leading to a merger fraction $f_m=0.11\pm0.04$. There is no evidence of redshift evolution of $f_m$ even at the highest redshift considered, thus extending well into the epoch of reionization the constant trend seen previously at $z\lesssim 6$. Furthermore, we investigate any potential redshift dependent differences in the specific star formation rates between mergers and non-mergers. Our analysis reveals no significant correlation in this regard, with deviations in the studied redshift range typically falling within $(1-1.5)\sigma$ from the null hypotesis that can be attributed to sample variance and measurement errors. Finally, we also demonstrate that the classification of a merging system is robust with respect to the observed (and equivalently rest-frame) wavelength of the high-quality JWST broad-band images used. This preliminary study highlights the potential for progress in quantifying galaxy assembly through mergers during the epoch of reionization, with significant sample size growth expected from upcoming large JWST infrared imaging datasets. 
\end{abstract}


%%%%%%%%%%%%%%%%%%%%%%%%%%%%%%%%%%%%%%%%%%%%%%%%%%

%%%%%%%%%%%%%%%%% BODY OF PAPER %%%%%%%%%%%%%%%%%%

\section{Introduction}\label{sec:introduction}

Interactions between galaxies critically influence their physical properties and play a core role in their mass accumulation and growth, driving morphological transformations and often resulting in mergers (\citealt{Toomre_1972}). The role of interactions and mergers is pivotal to our understanding of galaxy evolution, including how their impact may change over cosmic time. One key measurement is the fraction of the galaxy population undergoing a merger, denoted as merger fraction ($f_m$), and whether there is any dependence with redshift or on galaxy properties, such as stellar mass or luminosity.\\
\indent Direct imaging studies of the galaxy merger fraction broadly fall into two main methodologies; (i) examining the frequency of galaxies in close proximity to estimate upcoming mergers (“close pair identification method”, e.g., \citealt{Barnes_1988, LeFevrè_2000, Patton_2002,Schmidt_2013}) and (ii) measuring the morphology of individual sources to identify ongoing/recently completed mergers (“morphological parameters method”, e.g., \citealt{Conselice_2003, Lotz_2008, Conselice_2009a, Conselice_2009b}).\\
\indent Over the last two decades, investigations beyond the local Universe have predominantly utilized Hubble Space Telescope (HST) imaging. \citet{Lotz_2006} used the All-Wavelength Extended Groth Strip International Survey (AEGIS) to study galaxy morphologies and merger frequencies over the redshift range $0.2<z<1.2$ and determined a merger fraction consistent with no redshift evolution; measuring $f_m = 0.10\pm0.02$ using a parent sample with $I_{F814W}<25.0$.
This was supported by \citet{Conselice_2009a}, who identified a consistent merger fraction of $f_m\sim0.10$ for galaxies between redshifts $0.6<z<1.2$ and with stellar mass $M_*>10^{10}M_\odot$, using both Extended Groth Strip (EGS) (\citealt{Davis_2007}) and The Cosmic Evolution Survey (COSMOS) (\citealt{Scoville_2007}) surveys.\\
\indent In contrast, other studies focusing on massive galaxies ($M_*>10^{10}M_\odot$) lying on the star-forming main sequence (MS) at $0.2<z<2.0$ identified an increase in the merger fraction with increasing redshift; from $f_m\sim0.03$ at $z\sim0.5$ to $f_m\sim0.13$ at $z\sim1.5$ \citep{Speagle_2014,Cibinel_2019}. Additionally, these investigations found a dependence of the merger fraction on galaxy properties, with the likelihood of galaxies merging depending significantly on their distance from the MS. Analyzing massive galaxies with $M_* \geq 10^{10.8} M_{\odot}$ selected from the Ks-band catalogs of UltraVISTA/COSMOS and CANDELS/3DHST, \citet{Man_2016} observed an increasing trend in the merger fraction with redshift for $z < 0.65$. The merger fraction then stabilizes at approximately $f_m \sim 0.10$ up to a redshift of $z \sim 2.4$.\\
\indent \citet{Kim_2021} also determined a redshift evolution in the merger fraction in a mass-selected galaxy sample ($9.0<\log(M_*/M_{\odot})<11.5$) in the North Ecliptic Pole-Wide field using the morphological parameters of the candidates to detect mergers. Within their sample, they obtained a merger fraction of $f_m\in(0.10,0.20)$ at $z<0.6$, consistent with \citet{Conselice_2009a}, with a marginally higher fraction of $f_m\sim0.20$ for galaxies in the redshift range $4.0<z<6.0$, consistent with the high mass sample ($10^{9}M_{\odot}$) from \citet{Conselice_2009b} at the same epoch and with the same merger identification method.\\
\indent Moreover, \citet{Ventou_2017,Ventou_2019} examined the evolution of the merger fraction out to $z=6$ using the close pair identification method with data collected from MUSE (the Multi-Unit Spectroscopic Explorer) and determined that the evolution of the merger fraction is best described by two distinct behaviors. In the local Universe, they observed a significant decline in the merger fraction with cosmic time reaching the lower limit of $f_m=0.009\pm0.02$ at redshift $z=0.05$ \citep{dePropris_2007}. However, for redshift greater than $z\sim1$, the major merger fraction remained approximately constant at a value of around $f_m\sim0.1$. This fraction is determined by including the total number of galaxies in the sample, the number of close pairs consistent with major mergers (i.e. 1:6 mass ratio between secondary and primary galaxies) and those without companions due to spatial resolution limits (\citealt{Ventou_2017}). These findings are not only consistent with observational measurements, such as those in \citet{Conselice_2009a}, but also with simulations, including ILLUSTRIS \citep{Snyder_2017} and EMERGE \citep{O'Leary_2021}, especially when considering a sample of galaxies with a stellar mass $M_*\geq10^{9.5}M_{\odot}$. Importantly, none of these studies suggest an increase in the merger fraction beyond $z>2$.\\
\indent In contrast, a result indicating a higher merger fraction in the redshift range $4.4\leq z < 4.6$ and $5.1\leq z < 5.9$ for primary galaxies in the mass range of $9\lesssim \log(M_*/M_{\odot})\lesssim11$ was found in \cite{Romano_2021}. The study involved analyzing the [CII] emission line from a sample of $118$ star-forming galaxies to obtain morpho-kinematic information from the ALMA Large Program to INvestigate [CII] at Early times (ALPINE) survey (\citealt{Bèthermin_2020,Faisst_2020,LeFèvre_2020}). A major merger fraction of $f_m=0.44^{+0.11}_{-0.16}$ at $z\sim4.5$ and $f_m=0.34^{+0.10}_{-0.13}$ at $z\sim5.5$ is reported, consistent with the morphological study conducted by \cite{Conselice_2009b}, which found a merger fraction of $f_m = 0.23 \pm 0.05$ at $z= 3.8$ and $f_m = 0.19^{+0.05}_{-0.06}$ at $z= 4.8$ for galaxies with a stellar mass $M_* > 10^{9-10} M_{\odot}$.\\
\indent Significantly, investigations into the merger fraction using conventional methodologies, primarily relying on ground-based observatories and the Hubble Space Telescope, have been confined to redshifts $z<6$ due to the limited availability of high-resolution deep near-infrared (NIR) imaging and wavelength filters crucial for directly examining the rest-optical structures of high-redshift galaxies. However, with the advent of the James Webb Space Telescope (JWST), we can now expand the study of morphological characteristics and parameters of galaxies to higher redshifts, leveraging the enhanced sensitivity of the Near Infrared Camera (NIRCam) instrument. Recent studies have shown that the morphological parameters and the overall morphology of galaxies at the time of reionization ($z>7$) do not vary significantly between optical and UV rest-frame observations (e.g., \citealt{Treu_2023,Vulcani_2023,Tohill_2024}). This has been made possible thanks to the enhanced resolution and broader wavelength range provided by the JWST NIRCam observational filters compared to those of the HST.\\
\indent Notably in recent years, the research to understand galaxy mergers has transcended standard methodologies of identifying mergers by employing random forest (RF) and Machine Learning classifications trained on simulated JWST images from CEERS and the IllustrisTNG simulation, achieving accuracies of approximately $60\%$ \citep{Gomez_2019, Snyder_2019, Rose_2023}. These alternative methods have been utilized to measure the merger fraction and observe redshift evolution, revealing a transition from $f_m\sim 0.03$ at $z\sim 0.7$ to $f_m\sim0.40$ at $z\sim3.7$ \citep{Rose_2023}. This highlights significant deviations from studies using standard methods and underscores that the identification method of galaxy mergers at high redshifts holds the potential to introduce substantial systematic effects.\\
\indent In this study we conduct a comprehensive study of galaxy mergers in the epoch of reionization by leveraging the deep NIRCam imaging from the GLASS-JWST ERS program (\citealt{Treu_GLASS_2022}), UNCOVER JWST-GO-2561 (\citealt{Bezanson_2022}) and DDT-2756 (PI Wenlei Chen). 
These new observations provide us with the opportunity to extend the morphological studies out to redshift $z\sim 9$ (lookback time of about 13.3 Gyr), and offer a more extensive dataset in comparison to prior studies that relied on HST data. Our study focuses on the low-magnification regions in the outskirts of galaxy cluster Abell 2744, where we employ morphological statistical parameters to assess the prevalence of merger systems across a wide range of redshifts. Our objective is to extend the investigation of galaxy mergers to the high redshift, a domain that has not been thoroughly explored before, encompassing both bright and faint galaxies out to $z\sim9$.\\
\indent The structure of this paper is organized as follows: Sec.\ref{sec:data selection} provides a comprehensive overview of the data selection process for the galaxy candidates. 
In Sec.\ref{sec:analysis}, we outline the galaxy morphological parameters and merger criteria that we will employ to identify interacting systems. 
In Sec.\ref{sec: results and discussion} we determine the merger fraction and then examine and discuss any redshift evolution within our sample and whether there are any galaxy characteristics differences between mergers and non-mergers (e.g. the sSFR).   
Finally, in Sec.\ref{sec:summary}, we provide a summary of our discoveries. In App.\ref{app: wavelength dependency study}, we present a study exploring the influence of observation filter wavelengths on the analysis of morphological parameters.\\
\indent In this work we assume, when relevant, a standard cosmology with $H_{0}=70$ $\rm km\ s^{-1}Mpc^{-1}$, $\Omega_{\rm m} = 0.3$, $\Omega_\Lambda = 0.7$. and a \cite{Kroupa_2022} Initial Mass Function (IMF). Magnitudes are in the AB system \citep{1983ApJ...266..713O}.

\section{Observational data and sample selection}\label{sec:data selection}

\begin{table*}
\centering
\begin{tabular}{ccccc}
    \hline \hline
    \multicolumn{5}{c}{NIRCam observations} \\
    \hline
    \multicolumn{1}{c}{$z-bin$} & \multicolumn{1}{c}{$N_g$ } & \multicolumn{1}{c}{$\overline{z}$} & \multicolumn{1}{c}{M$_\mathrm{U}$} & \multicolumn{1}{c}{$\log_{10}(M_*/M_\odot)$}\\ 
    (1) & (2) & (3) & (4) & (5)\\ 
    \hline \hline
    $4.0\leq z<5.0$ & $845(\textbf{319})$ & $4.50$ & $[-26.6,-17.9]$ & $[6.9,11.4]$\\ 
    $5.0\leq z<6.0$ & $1333(\textbf{186})$ & $5.34$ & $[-24.7,-18.3]$ & $[7.2,10.9]$\\ 
    $6.0\leq z<7.0$ & $386(\textbf{107})$ & $6.47$ & $[-23.1,-18.5]$ & $[7.3,9.9]$\\ 
    $7.0\leq z<8.0$ & $414(\textbf{30})$ & $7.47$ & $[-23.6,-18.7]$ & $[7.6,10.7]$\\ 
    $8.0\leq z<9.0$ & $250(\textbf{33})$ & $8.50$ & $[-25.6,-19.3]$ & $[7.5,12.4]$\\  
    \hline \hline
\end{tabular}
\caption{Summary of the galaxies used in this work for mergers identification, obtained using JWST NIRCam data from the GLASS Collaboration. The sources are selected as discussed in Sec.\ref{sec:data selection}. (1) Redshift bin. (2) Number of galaxies detected.. (3) Mean redshift of the subsample. (4) Absolute U Magnitude range of the subsample. (5) Stellar mass range. Quantities in columns (3), (4) and (5) are referred to the subsamples used in this work in \textbf{bold} (2). Both distributions of magnitude and stellar mass in respect to redshift are presesented in Fig.~\ref{fig: magnitude-mass vs z - hres - mu2.0}.}
\label{tab:candidates}
\end{table*}

\begin{figure}
\includegraphics[angle=0,width=1.\linewidth]{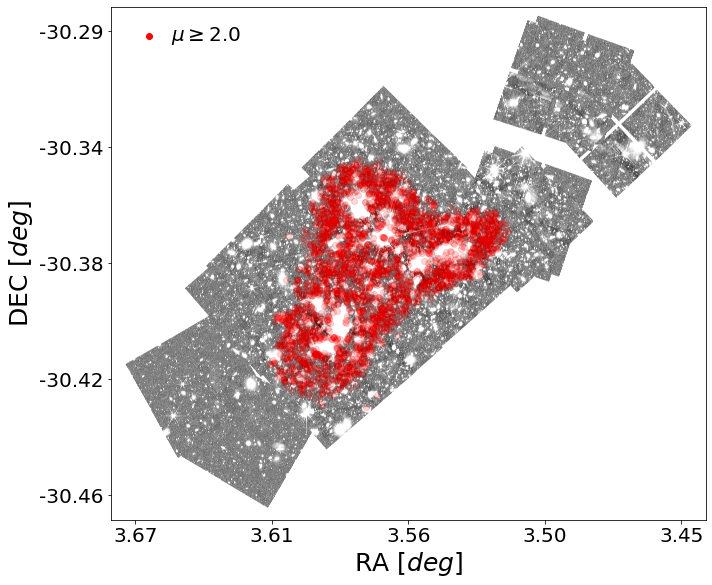}
\caption{NIRCam scientific map in the F150W filter. The red region represents the area where the magnification  is $\mu\geq2.0$. The galaxies detected in this region of the survey are excluded from the parent sample.}
\label{fig:magnification cut}
\end{figure}

The NIRCam imaging data used in this study were obtained through three public programs focused on the foreground galaxy cluster Abell 2744 and its immediate surroundings: (i) GLASS JWST-ERS-1324 \citep{Treu_GLASS_2022}, (ii) UNCOVER JWST-GO-2561 \citep{Bezanson_2022}, and (iii) the Director’s Discretionary Time Program 2756 (PI Wenlei Chen). We use the publicly available galaxy and photometric catalogs for the combined footprint of these three programs, as provided by \citet{Merlin_2022,Paris_2023} with the photometric redshifts estimated by fitting the full HST+JWST photometry to \citet{Bruzual_2003} templates by means of the \textsc{zphot} code (\citealt{Fontana_2000}), using the same technique adopted in \citet{Santini_2023}. Additionally, we utilize the high-resolution reduction (hereafter referred to as \texttt{hres}, with $1$ pixel $= 0.02"$) of the F150W and F200W NIRCam filters (with filter windows of 1300-1700 \AA\ and 1700-2300 \AA\ respectively) by \citet{Brammer_2023} over the entire survey area, as illustrated in Fig.\ref{fig:magnification cut}.\\
\indent Because we focus on a high redshift sample of galaxies, we restrict the \citet{Paris_2023} source catalog to candidates with a photometric redshift in the range of $4\leq z<9$. We set an upper bound at $z=9$ due to the low number of candidates at such a high redshift, limiting any statistical constraints we can place on our population analysis. This provides an initial catalogue of 3228 galaxies. To ensure we can robustly measure the morphological parameters, we imposed both a signal-to-noise ratio (SNR) and a star class requirements, demanding that candidate galaxies in the observed NIRCam imaging bands exhibit a total SNR exceeding $SNR = 7.0$ in each band and that the value of the stellarity index \texttt{class}$\_$\texttt{star} measured by the \texttt{SExtractor} code (\citealt{Bertin_1996}) is not consistent with that of a point source, i.e., we require that \texttt{class}$\_$\texttt{star}$<0.9$.\\
\indent Additionally, for our morphological analysis we aim to ensure that the parameters we measure reflect the true shape of the galaxies and are not deformed by gravitational lensing, which is important in our study because our NIRCam footprint includes the Abell 2744 galaxy cluster. We considered the impact of gravitational lensing and opted to exclude regions with excessively high magnification coefficients. Galaxies were deemed suitable if their median magnification, computed by \citet{Bergamini_2023}, fell below $\mu < 2.0$. Fig.\ref{fig:magnification cut} illustrates the region excluded by this criterion, highlighted in red.\\ 
\indent Finally, to ensure the completeness of our sample, we implement an apparent magnitude cutoff, making a conservative assumption using the limiting AB magnitude at $5\sigma$ for NIRCam in the F150W band (rest-frame UV 1500-3000\AA\ for our sample), which was set at $m_{AB}= 28.87$. Absolute UV magnitudes at 1500\AA~and stellar masses were calculated by fitting the photometry with \citet{Bruzual_2003} templates and assuming a delayed exponentially declining SFH, as done in \citet{Santini_2023}.\\ 
\indent Tab.\ref{tab:candidates} shows the number of candidates detected by NIRCam (total: 3228) in each redshift bin and the number of targets after the application of the selection criteria used in this work (total: 675). \\
\indent Given that the angular-diameter distance is relatively constant at high z, to conduct morphological parameters estimations for each galaxy we consider a cutout from the composite high-resolution map (with $1$ pixel $= 0.02"$) of $110$px on a side corresponding to $2.2"$ on a side (our science images are drizzled on a $0.02"$/pixel scale).

\section{Analysis}\label{sec:analysis}

\subsection{Morphological Parameter Definition\label{subsec:morph params defs}}

In this work we employ three widely recognized quantitative morphological statistics, employed for characterization of mergers (\citealt{Conselice_2014}): the Gini coefficient ($G$), the second-order moment of brightness ($M_{20}$), and the asymmetry ($A$). The definitions are given below for convenience of the reader and to set the notation\footnote{The morphological parameters were measured using \texttt{JWSTmorph}, a publicly accessible code available on the GitHub repository:~\url{https://github.com/Anthony96/JWSTmorph.git}}.\\
\indent The Gini coefficient, proposed as a morphological parameter by \citet{Abraham_2003}, quantifies the inequality in the distribution of pixel intensities within a galaxy image. It is computed as:
\begin{equation}
    G=\frac{1}{\bar{X} n(n-1)} \sum_{i}^{n}(2 i-n-1) X_{i}
\end{equation}
where $X_i$ represents the intensity of the $i^\mathrm{th}$ pixel, $n$ is the total number of pixels assigned to the galaxy from the segmentation map, and $\bar{X}$ is the mean intensity. By definition, this parameter falls within the range of $[0,1]$, where a value of 0 implies that the galaxy exhibits a uniform distribution in terms of intensity, while a value of 1 indicates that one pixel possesses all the flux. For galaxy mergers, it is anticipated that this morphological parameter will be higher compared to non-mergers. This expectation arises from the fact that the process of galaxy mergers often results in a more concentrated distribution of light within the merged system. This concentration occurs due to interactions between galaxies and the central concentration of stellar material, contributing to a higher value of this parameter (e.g., \citealt{Lotz_2008}).\\
\indent The second-order moment of brightness $M_{20}$, introduced by \citet{Lotz_2004}, measures the compactness and concentration of the brightest $20\%$ of a galaxy's light. It is defined as:
\begin{equation}
    M_{20}=\log _{10}\left(\frac{\sum_{i} M_{i}}{M_{t o t}}\right), \text { with } \sum_{i} f_{i}<0.2 f_{\text{tot}} \\
\end{equation}
where $f_{\text{tot}}$ is the total flux of the galaxy pixels identified by the segmentation map, $M_i$ is the second-order moment of brightness for each pixel, and $M_{\text{tot}}=\sum_{i}^{n} M_{i}=\sum_{i}^{n} f_{i}\left[\left(x_{i}-x_{c}\right)^{2}+\left(y_{i}-y_{c}\right)^{2}\right]$, with $f_i$ being the single pixel intensities, $x_{i}$ and $y_{i}$ are the pixel coordinates, while $x_{c}$ and $y_{c}$ correspond to the galaxy center where $\mathrm{M}_{\text {tot }}$ is minimized. Typical values for this parameter are typically found in the range of $[-3,0]$. Increasing values within this range are correlated with a greater number of off-centered bright features associated with the galaxy under study. The merger process tends to make the central region of the merging galaxies more concentrated and brighter. Consequently, this concentration has a significant impact on the $M_{20}$ parameter. Non-merging galaxies, on the other hand, typically exhibit a smoother and less concentrated distribution of bright pixels, resulting in lower $M_{20}$ values. These galaxies may lack the distinct, compact, and bright regions that are characteristic of merging systems, $M_{20}$ is expected to be higher for galaxy mergers compared to non-mergers.\\
\indent The asymmetry parameter ($A$), proposed by \citet{Abraham_1996} and \citet{Conselice_2000}, is calculated as:
\begin{equation}
    A=\frac{\Sigma\left|I -I_{\pi}\right|}{\Sigma I}-A_{bkg}
\end{equation}
where $I$ is the original cutout image, $I_{\pi}$ is the image rotated by 180 degrees (or $\pi$ radians), and $A_{\text{bkg}}$ denotes the asymmetry of the background.  
By construction, the asymmetry parameter tends to be higher for galaxy mergers compared to non-mergers. This is because mergers usually disrupt structures, introduce asymmetry and irregularities into a galaxy's appearance.\\
\indent These morphological parameters are measured using \texttt{hres} image cutouts of the candidates selected in Sec.\ref{sec:data selection} using \texttt{JWSTmoprh}. The segmentation maps, which are used to assign the pixels to a galaxy, are derived as in \citet{Treu_GLASS_2022} through the \textit{photutils} package\footnote{\url{https://photutils.readthedocs.io/en/stable/citation.html}} \citep{Bradley_2023}, using a threshold flux for detection of $2\sigma$ above the background. \\

\subsection{Identifying Mergers}\label{subsec: merger identification}

\begin{figure*}
\includegraphics[width=1.0\linewidth]{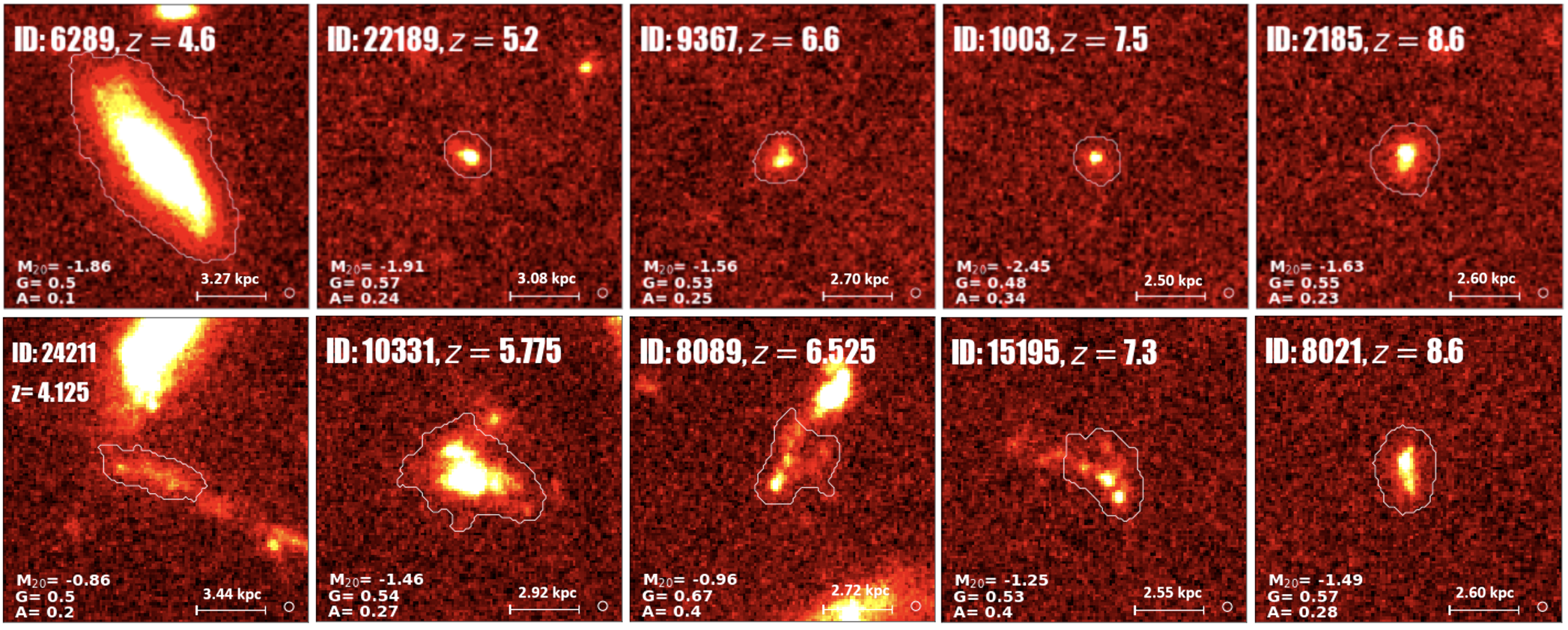}
\caption{ Illustrative representations from various redshift ranges, featuring extracted cutouts alongside corresponding morphological metrics, ID and photometric redshift. These snapshots are derived from a composite \texttt{hres} map ($20$mas per pixel), integrating data from F150W and F200W bands. The upper row highlights non-merging galaxies, while the lower row presents merger candidates. Each cutout stamp is $2.2\arcsec$ on a side, while the horizontal bar is $0.5 \arcsec$ wide and the white contour represents the $2\sigma$ segmentation map boundary.}
\label{fig: merger vs nonmerger - hres - mu2.0}
\end{figure*}

The morphological parameters described above can be used to classify whether a galaxy is a merger, and \cite{Conselice_2003} and \citet{Lotz_2008} set out merger criteria using the two equations: 
\begin{eqnarray}\label{eq: gini classical merger criteria}
    f(G,M_{20}) = G + 0.14M_{20} > 0.33\\
\label{eq: asimmetry classical merger criteria}
    A \geq 0.35
\end{eqnarray}
These two equations have been derived from HST observations and have been validated for lower redshifts, extending only up to  $z\sim1.2$. It is worth noting that this redshift range falls considerably short of our intended investigation, which targets galaxies at $z\geq4$. Nevertheless, recent research \citep{Treu_2023,Vulcani_2023} has revealed minimal variations in the underlying parameters with respect to redshift. In light of these findings, we proceed to employ these equations in our examination of our galaxy sample. To facilitate classification, we introduce two distinct samples:\\
\indent $\bullet$ Silver: galaxies that satisfy the first criteria Eq.\ref{eq: gini classical merger criteria} are classified as members of the Silver sample.\\
\indent $\bullet$ Gold: galaxies that satisfy both criteria equations Eq.\ref{eq: gini classical merger criteria} and Eq.\ref{eq: asimmetry classical merger criteria} are classified as members of the Gold sample.\\
\indent Note that according to the definition above, the candidates classified as mergers in the Gold sample are a subset of those in the Silver sample. The difference between these two subsamples is in a requirement on the asymmetry parameter $A$. Physically, this means that in addition to considering the flux and brightness distribution from $G$ and $M_{20}$, we also account for the pixel non-uniformity caused by the disruptive merging process, indicating a more dispersed galaxy.\\
\indent By employing these classification criteria, we are able to categorize mergers based on their adherence to specific conditions, enabling a more systematic analysis of merger populations in the field of astrophysics \citep{Lotz_2008}.\\
\indent In Fig.\ref{fig: merger vs nonmerger - hres - mu2.0}, we offer a representative example of galaxies categorized as mergers and non-mergers in each redshift bin. The upper row showcases a non-merger candidate, while the lower row presents a merger candidate, the associated morphological parameters are presented as a reference.\\

\section{Results and Discussion}\label{sec: results and discussion}

\subsection{Mergers characterization}\label{subsec: mergers characterization}

\begin{figure*}
\includegraphics[width=1.\linewidth]{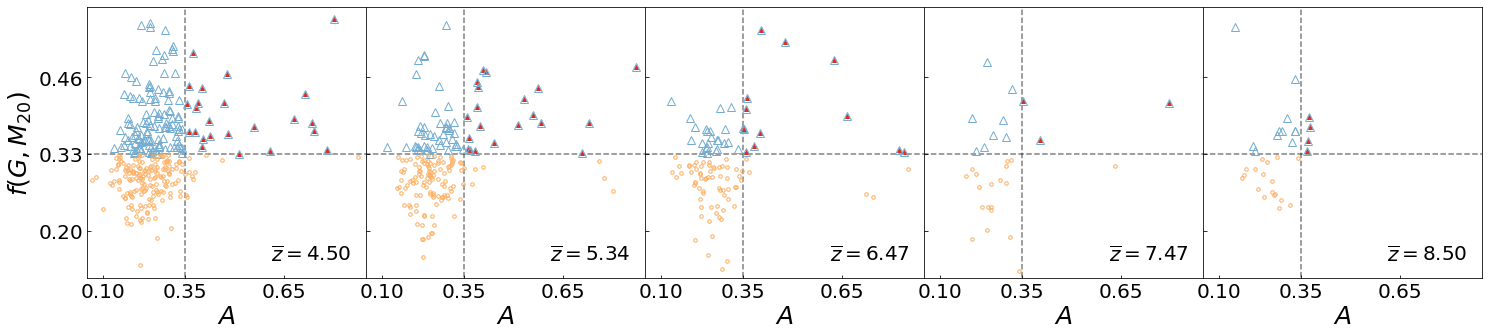}
\caption{Visual representation of the classification method adopted in this study for five different z-bins (see Sec.\ref{subsec: merger identification}). Galaxies located above the horizontal dashed line (representing Eq.\ref{eq: gini classical merger criteria}) are labelled with blue triangles and are classified in the Silver sample. Among these, the subset to the right of the vertical dashed line (representing Eq.\ref{eq: asimmetry classical merger criteria}) is highlighted in red, representing the galaxies in the Gold sample. We used high-resolution reduction of the F150W and F200W NIRCam imaging (see Sec.\ref{sec:data selection}).}
\label{fig: GM20 vs A - hres - mu2.0}
\end{figure*}

\begin{figure}
\includegraphics[width=1.\linewidth]{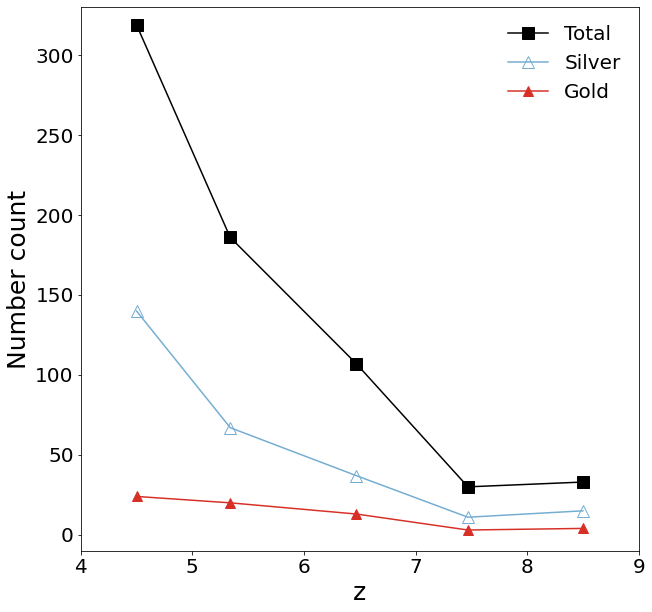}
\caption{Demography for the galaxy (merger) samples versus redshift. Points correspond to those in Fig.\ref{fig: GM20 vs A - hres - mu2.0}. Black squares indicate the total number of candidates considered in each redshift bin (bold number in column (2) in Tab.\ref{tab:candidates}). Blue triangles represent galaxies in the Silver sample, while red triangles denote those classified in the Gold sample.}
\label{fig: number G_GS_tot - hres - mu2.0}
\end{figure}

\begin{figure*}
\includegraphics[width=1.0\linewidth]{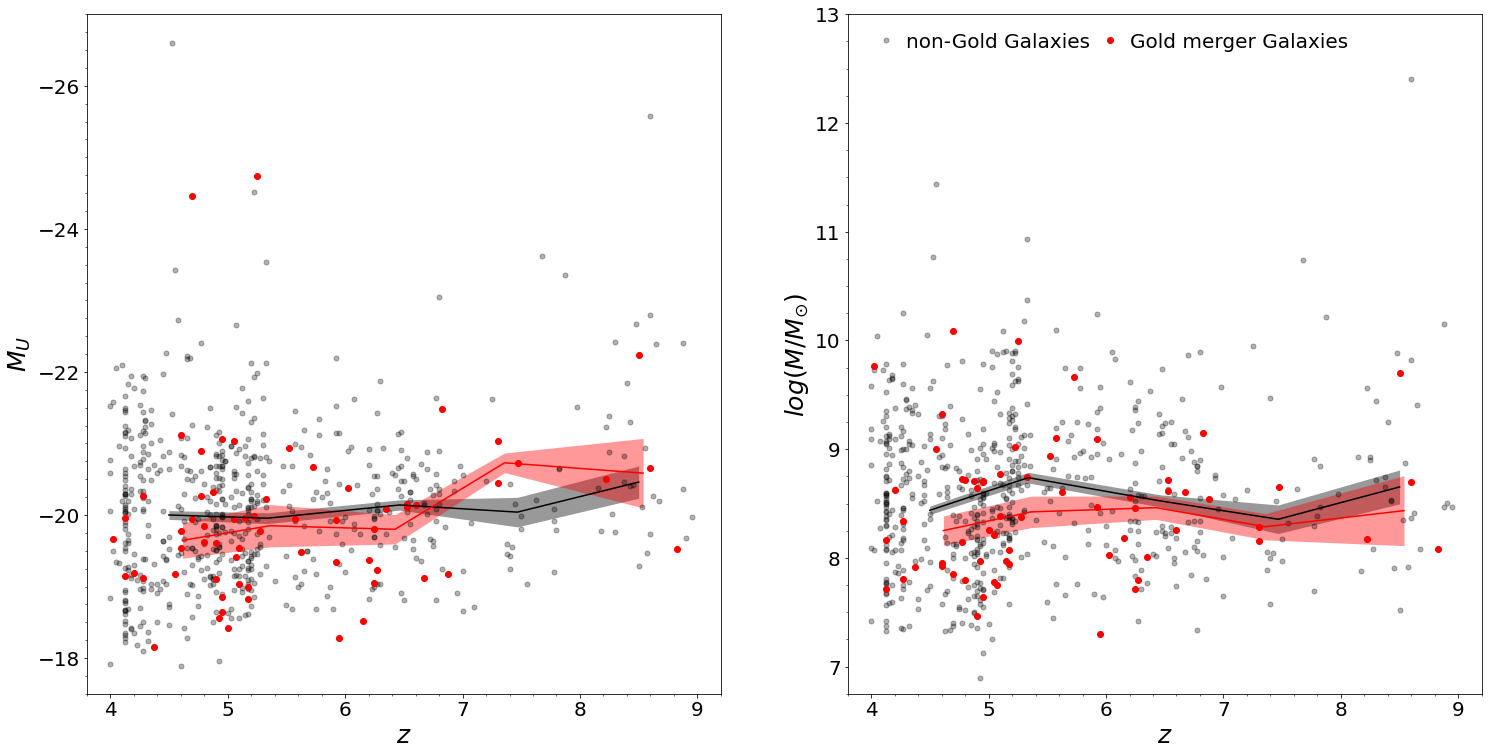}
\caption{Rest frame U-magnitude (left) and stellar mass (right) as a function of photometric redshift for the Gold sample (red) and for all other, non-Gold galaxies (grey). Solid lines represent the corresponding median evolution in respect to the mean redshift in each bin} of the two different samples of mergers and non-mergers. Shaded regions represent the standard error corresponding to these median values.
\label{fig: magnitude-mass vs z - hres - mu2.0}
\end{figure*}

\begin{figure}
\includegraphics[width=1.0\linewidth]{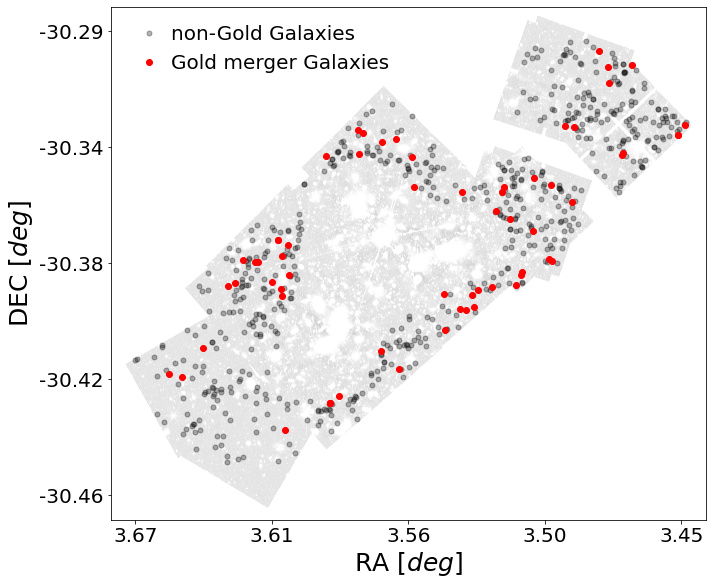}
\caption{The Gold sample and non-Gold galaxies obtained in all the different redshift bins. Note that the central region is not populated as we excluded regions with magnification $\mu \geq 2.0$, Fig.\ref{fig:magnification cut}.} 
\label{fig:RA DEC mu2 cut merger}
\end{figure}

We measure the morphological parameters for the 675 galaxies in our parent sample, and use these to define a Silver and Gold sub-sample of mergers based on our merger criteria. We measure these parameters from the \texttt{hres} imaging, and note that we find consistent values for the structural parameters and merger classifications when using alternative filters (see App.\ref{app: wavelength dependency study}).
Our classification outcomes are visually depicted in Fig.\ref{fig: GM20 vs A - hres - mu2.0}, where the two dashed lines represent the threshold values derived from Eq.\ref{eq: gini classical merger criteria} and \ref{eq: asimmetry classical merger criteria}. Each data point on the graph corresponds to an individual galaxy from our sample, chosen based on the criteria detailed in Sec.\ref{sec:data selection} and summarized in Tab.\ref{tab:candidates}.\\
\indent Fig.\ref{fig: number G_GS_tot - hres - mu2.0} provides the respective count of galaxies falling into each region defined by the dashed lines in Fig.\ref{fig: GM20 vs A - hres - mu2.0}. The black squares represents the total sample of galaxies that meet the selection criteria outlined in Sec.\ref{sec:data selection}, the blue triangles designates the Silver sample, while the red triangles represents the Gold sample.\\
\indent To offer a more comprehensive characterization of the candidate sample selected for this study and the identified mergers, we additionally present two figures. Fig.\ref{fig: magnitude-mass vs z - hres - mu2.0} shows the galaxy absolute magnitude in U-band (left panel) and stellar mass (right panel) as a function of redshift for the Gold sample and the non-Gold galaxies (all the galaxies excluded from the Gold sample). It is apparent that the two distinct categories of galaxies exhibit a comparable range in terms of both magnitude and mass, so we can state that mergers are not typically brighter or more massive than isolated galaxies (to $2\sigma$) and this inference is valid across the whole redshift range of our analysis. Furthermore, for a spatial perspective and to better visualize the distribution of classified mergers, Fig.\ref{fig:RA DEC mu2 cut merger} displays the F150W image in (RA, DEC) coordinates. Overlaid on the image are galaxies classified as mergers (Gold sample) and those that do not fall into this category. This visualization is based on high-resolution data and covers the entire redshift range, specifically $z\in [4.0,9.0]$. From Fig.\ref{fig:RA DEC mu2 cut merger} we can see that there is no peak in the spatial distribution of the merger galaxies contained in the Gold sample; they are more or less uniformly distributed within the area enclosed in the foreground of the Abell 2744 cluster field.

\subsection{Merger fraction versus redshift}\label{subsec: mf vs z }

\begin{figure*}
\includegraphics[width=1.0\linewidth]{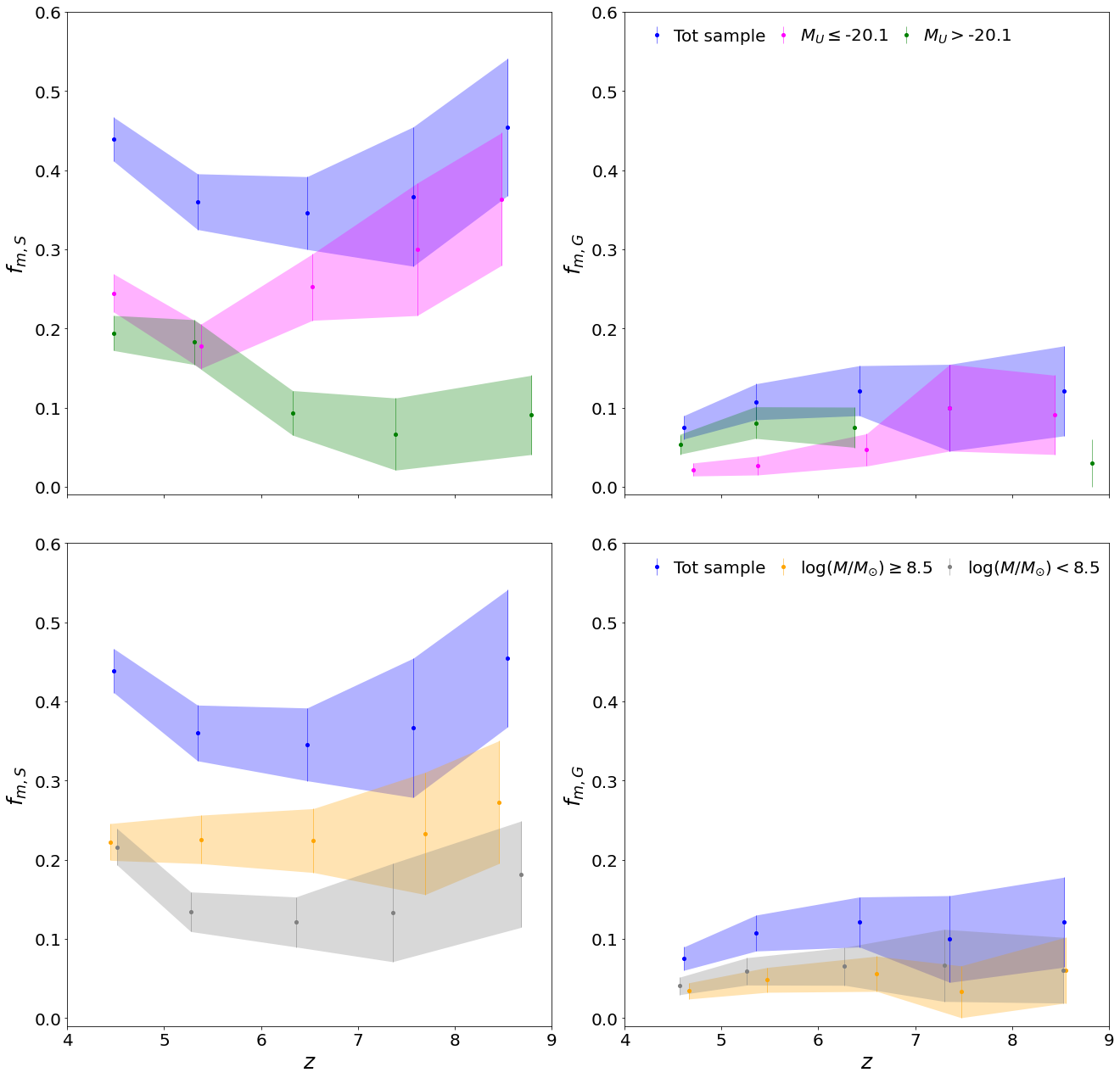}
\caption{Merger fraction of the Silver sample (left panels) and of the Gold sample  (right panels) as a function of redshift. The definition of the fraction follows Eq.\ref{eq: fm tot}. In the top panels, we show a further separation based on the rest frame magnitude, with bright galaxies classified as having U-band magnitudes below $M_{U}=-20.1$, and faint galaxies lying above this threshold. In the bottom panels, we separate galaxies into a high mass sample ($M_* > 10^{8.5} M_\odot$) and a low mass sample ($M_* < 10^{8.5} M_\odot$).}
\label{fig: fm_GS-G vs z - magnitude and mass cut - hres - mu2.0}
\end{figure*}

\begin{figure}
\includegraphics[width=1.0\linewidth]{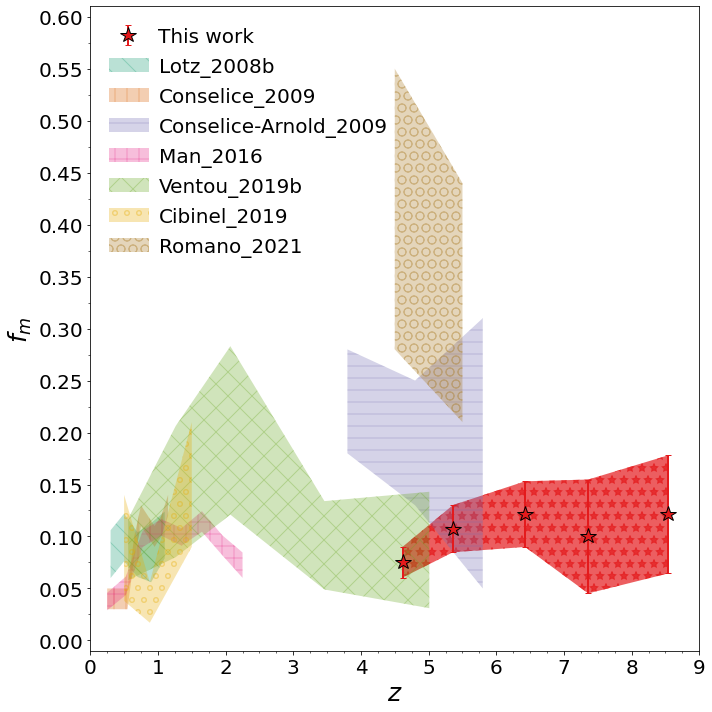}
\caption{Comparison of the merger fraction ($f_m$) evolution with respect to redshift in our work against literature measurements from: \citealt{Lotz_2006,Conselice_2009a,Conselice_2009b,Man_2016,Ventou_2019,Cibinel_2019,Romano_2021}. Note that each study implements their own catalogs, data selection, and methodologies for identifying mergers, leading to a non homogeneous comparison that may be affected substantially by systematic effects.}
\label{fig: fm_vs_z_comparison}
\end{figure}

In each redshift bin, we define the merger fraction $f_{m,i}$ as the number of mergers ($N_{m}$) over the total number of galaxies ($N_{\text{tot}}$) in the $i$-th redshift bin:
\begin{equation}\label{eq: fm tot}
f_{m,i}=\left(\frac{N_{m}}{N_{\text{tot}}}\right)_i
\end{equation}
\indent To explore the potential trend of merger fraction with magnitude, we implement a magnitude cutoff to divide the sample into bright and faint galaxies. We set a magnitude threshold at $M_{U}=-20.1$, the median of all galaxies presented in the left panel of Fig.\ref{fig: magnitude-mass vs z - hres - mu2.0}. We apply a similar approach when creating two sub-samples by introducing a stellar mass cutoff for the candidates, with a threshold of $\log(M/M_{\odot}) = 8.5$. This threshold corresponds to the median of all galaxies presented in the right panel of Fig.\ref{fig: magnitude-mass vs z - hres - mu2.0}.\\
\indent In Fig.\ref{fig: fm_GS-G vs z - magnitude and mass cut - hres - mu2.0}, we depict the merger fraction across the redshift range, considering both the Gold and Silver samples examined in this study. We present the merger fraction $f_m$ in the top two panels when considering the total sample of galaxies, alongside two sub-samples created by implementing a magnitude cutoff to distinguish them. The two lower panels show the same quantity, but this time, the sample is segregated based on stellar mass.\\
\indent The distinction in defining these two samples has a noticeable impact on the overall merger fraction across all the redshifts considered in our study. This effect is visually evident in Fig.\ref{fig: number G_GS_tot - hres - mu2.0} when we look at the number of galaxies in each group. When we consider our entire sample (blue regions in Fig.\ref{fig: fm_GS-G vs z - magnitude and mass cut - hres - mu2.0}) we observe minimal variation with redshift in the measured merger fraction with redshift for our Silver and Gold criteria. Specifically, for the Silver sample, the merger fraction is measured to be $f_m = 0.39\pm0.06$ when considering the entire redshift range. In contrast, the Gold sample exhibits a substantially lower of $f_m = 0.11\pm0.04$. The choice between the Silver and Gold samples has a significant impact on the derived merger fraction values. While the Silver sample, with its more permissive criteria, yields higher merger fractions across the redshift range, the Gold sample, which employs stricter criteria, results in consistently lower and more stable merger fractions.\\
\indent Earlier investigations (e.g. \citealt{Lotz_2008, Lotz_2006,Lin_2008,Conselice_2009a,Lòpez-Sanjuan_2009,Jogee_2009,Man_2016,Kim_2021}) on galaxies at $z\lesssim3$ consistently observed that the merger fraction tends to exhibit a relatively stable pattern around the value of $f_m\sim 0.10$, showing no significant overarching trend in relation to redshift. Studies conducted at higher redshifts (e.g.,\citealt{Ventou_2017,Ventou_2019}) also suggest that beyond a certain redshift threshold, likely below $z<1$, the merger fraction remains constant around a value of $f_m\sim0.10$ and does not display significant fluctuations. Our assessments of merger fractions based on Gold-criteria identification reveal no discernible redshift evolution, with an $f_m$ value consistent with findings from lower redshift investigations.\\
\indent Fig.\ref{fig: fm_vs_z_comparison} is a comparative plot to show how the redshift range $0<z<9$ is populated by the results obtained with different methods and data selections for studying the evolution of the merger fraction with redshift. In particular, \cite{Lotz_2006} conducted a morphological study on galaxies at low redshift $0.2 < z < 1.2$ with luminosity $L_{IR} > 10^{11}L_{\odot}$. In a similar redshift range and also using morphological analysis, \cite{Conselice_2009a} focused on galaxies with stellar mass $M_* > 10^{10}M_{\odot}$, extending the study to $4 < z < 6$ in \citet{Conselice_2009b}. Using a pair statistic approach from \cite{Man_2016} with the UltraVISTA/COSMOS Ks-band selected catalog, the authors measured the major merger fraction for galaxies with stellar mass $M_* \geq 10^{10.8}M_{\odot}$ within the redshift range $0.2 < z < 3$. Similarly, \cite{Cibinel_2019} used a combination of close pairs and morphological analysis in a sample of low redshift galaxies ($0.2 \leq z \leq 2.0$) with stellar mass exceeding $M_* > 10^{10}M_{\odot}$. \cite{Ventou_2019} analyzed close pairs from cosmological simulations in the redshift range $0.2 < z < 6$ for galaxies with stellar mass $M_* > 10^{9.5}M_{\odot}$. In \cite{Romano_2021}, a combination of morphological and kinematic measurements was employed, using emission lines of [CII] from a sample of galaxies with $4.4 < z < 5.9$ and stellar mass $9 \lesssim \log(M_*/M_{\odot}) \lesssim 11$. The results presented in this paper are shown in Fig.\ref{fig: fm_vs_z_comparison}, with a red star representing the Gold sample.\\
\indent Our main results for the merger fraction, considering the primary Gold sample, differ from some studies at intermediate redshift ranges ($4<z<6$) that derive a higher merging fraction, but use a different method for classification. Specifically, studies based on kinematic properties from emission spectra of galaxies, such as \cite{Romano_2021}, show a merger fraction of $f_m=0.44^{+0.11}_{-0.16}$ at $z\sim4.5$ and $f_m=0.34^{+0.10}_{-0.13}$ at $z\sim5.5$. Interestingly, our Silver sample results, where our selection criteria are less stringent (Sec.\ref{subsec: merger identification}), agree with these studies, leading to a merger fraction of $f_m\sim 0.39 \pm 0.06$ for intermediate redshift. Given the intrinsic uncertainties and limitations of morphological selection of merger candidates, a NIRSpec follow-up would be highly beneficial to understand the differences in the Gold versus Silver samples, and contribute to a  robust and accurate merger fraction measurement.\\
\indent Interestingly, we measure a merger fraction which is nearly redshift independent. To investigate the physical nature of this result, we can consider a simple model, where the merger fraction $f_m$  is expressed as a function of the merger rate:
\begin{equation}\label{eq:Gamma}
    f_m(z) = \Gamma_m(z) \cdot \tau_m(z)
\end{equation}
where $\Gamma_m \equiv dN_m/dt$ is the merger rate, representing the number of mergers per galaxy per unit of time. $\tau_m$ is the typical timescale in which a close pair will eventually merge into a single system, assuming that all close pairs will undergo a merging event. In studying the merger rate of dark matter halos, \cite{Fakhouri_2008} discuss how this quantity exhibits an almost universal behavior, showing a proportionality with the redshift of $\Gamma_{DMH} \propto (1+z)^{\beta}$ with $\beta=2-2.3$ up to high redshift $z \leq 7$. This has been derived also for galaxy merger events, finding the same redshift proportionality as for dark matter halos with $\beta=1.8$ out to $z \leq 3$ in \cite{Hopkins_2010} and $\beta = 2.43-2.87$ with $z \leq 8$ in \cite{RGomez_2015}. The expected increase in $\Gamma_m$ with increasing redshift from the dark matter halo dynamics is balanced by a decrease in $\tau_m$ as galaxies become smaller at higher redshift and thus the dynamical timescale decreases. This trend has been quantified by previous work  \citealt{Snyder_2017,Romano_2021}) which derive $\tau_m \propto (1+z)^{-2}$. As a result, the expression in Eq.~\ref{eq:Gamma} becomes approximately constant with redshift, in agreement with our results and previous measurements, including \cite{Conselice_2009a, Ventou_2017, Mantha_2018}.\\

\subsection{ sSFR of interacting and non-interacting systems}\label{subsec: interacting systems}

\begin{figure}
\includegraphics[width=1.\linewidth]{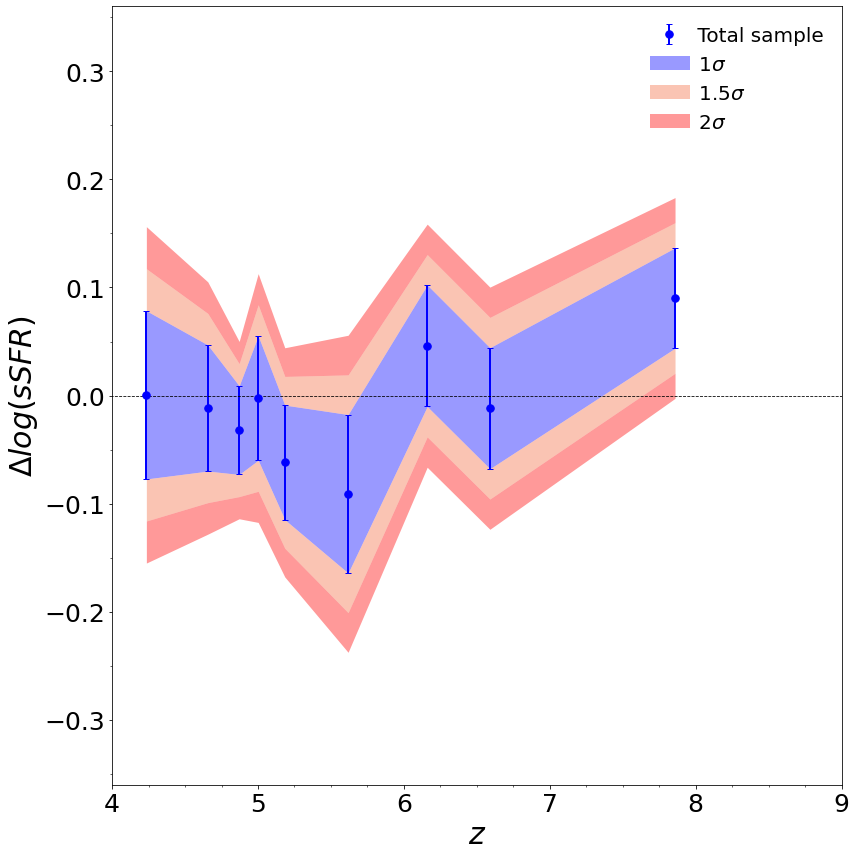}
\caption{Excess specific star formation rate (sSFR) evolution as a function of redshift for both the complete set of Gold sample merger and non-Gold galaxies (see Eq.\ref{eq: sSFR}). The shaded areas indicate uncertainty ranges of $\sigma = (1, 1.5, 2)$. The measurement is statistically consistent with no redshift evolution.}
\label{fig: sSFR vs z - hres - mu2.0}
\end{figure}

In this section we aim to explore potential variations in the impact of galaxy mergers on specific star formation activity across distinct redshifts.\\
\indent To derive the star formation rates, we use \texttt{BAGPIPES} \citep{Carnall_2018} to fit the photometries of the galaxies. \texttt{BAGPIPES}' model galaxy spectra were generated with a \citet{Kroupa_2022} initial mass function, a \citet{Calzetti_2000} dust attenuation law, a BPASS \citep[V2.2.1][]{Eldridge_2009} stellar population, and nebular emission lines from \texttt{CLOUDY} photoionization code \citep{Ferland_2017}. For the fit, we set the metallicity range to [0,1.2] $Z_\odot$, the dust attenuation $A_V$ to [0,3] mag, the logarithm of the ionization parameter $U$ to [-4,-1], and the logarithm of the total mass of stars formed in solar mass to [6,13]. We assumed two star formation histories: log-normal and non-parametric forms \citep{Iyer_2019}.\\
\indent We utilize a metric representing the excess of specific star formation rate (sSFR) for galaxies classified as mergers (Gold sample)in comparison to non-mergers (non Gold sample). This excess is calculated in the i-th redshift bin using the equation:
\begin{equation}\label{eq: sSFR}
\Delta \log(\text{sSFR}_i) = \log(\frac{<\text{sSFR}_{\text{merger}}>_i}{<\text{sSFR}_{\text{non-merger}}>_i})
\end{equation}
the associated standard uncertainty from error propagation is used. To present results across our redshift range $4\leq z<9$ we decided to partition the sample of objects to ensure a roughly uniform count of mergers in each redshift bin.\\
\indent According to the definition, values of $\Delta \log($sSFR$)>0$ indicate that galaxies identified as mergers exhibit a higher rate of star formation per unit mass compared to galaxies classified as non-mergers. Therefore, these measurements would suggest that those candidates are currently experiencing heightened levels of star formation relative to a reference or baseline. Conversely, data points falling below the zero baseline ($\Delta \log($sSFR$)<0$) indicate merger events where the sSFR is lower than what we observe in galaxies classified as non-mergers.\\
\indent In Fig.\ref{fig: sSFR vs z - hres - mu2.0} we show the evolution of the estimator defined in Eq.\ref{eq: sSFR} with respect to redshift. The fluctuations around the zero line fall within the $1\sigma$ range for the majority of the data points, while the rest fall within the $2\sigma$ range. These findings suggest that there is no strong indication of any notable increase or decrease in the sSFR evolution for mergers compared to non-mergers.
 
\section{Summary and perspectives}\label{sec:summary}

In this study, we focus on high-z galaxies in the outskirts (low magnification regions) of the foreground galaxy cluster Abell 2744, observed as part of the GLASS-JWST program ERS-1324  \citep{Treu_GLASS_2022}, UNCOVER JWST-GO-2561 \citep{Bezanson_2022} and DDT-2756 (PI Wenlei Chen).\\
\indent Our main objective was to use high-resolution ($1\text{px} = 0.02"$) F150W and F200W imaging to conduct an in-depth analysis of the evolution of the merger fraction $f_m$ with respect to redshift, extending from $z=4$ up to $z\simeq9$. We accomplished this by investigating various subsamples, which were derived based on magnitude and mass thresholds. Our key findings are as follows:\\
\indent $\bullet$ We investigated morphological parameters for high-redshift galaxies, categorizing them into two distinct merger subsamples: Silver and Gold. We employed the criteria outlined in Eq.\ref{eq: gini classical merger criteria} and Eq.\ref{eq: asimmetry classical merger criteria}, as defined in prior research on this topic. Our results are depicted in Fig.\ref{fig: GM20 vs A - hres - mu2.0}, illustrating the distribution of morphological parameters, and in Fig.\ref{fig: number G_GS_tot - hres - mu2.0}, which provides a visual representation of the number of candidates in each sample, correlated with the mean redshift of the candidates within the selected bins.\\
\indent $\bullet$ To obtain a more comprehensive understanding of the merger population in comparison to non-interacting galaxies, we have provided visual representations of the magnitude and mass distributions in Fig.\ref{fig: magnitude-mass vs z - hres - mu2.0}. These distributions have been segregated between candidates identified as Gold mergers and all galaxies included in the study, adhering to the selection criteria outlined in Sec.\ref{sec:data selection}. It is worth noting that no distinct trends have emerged from our analysis. We have observed that mergers are present across the entire range of magnitudes and masses at different redshifts. Notably, the distributions of these two categories exhibit similarities, as indicated by the medians of both magnitude and mass.\\
\indent $\bullet$ Fig.\ref{fig: fm_GS-G vs z - magnitude and mass cut - hres - mu2.0} displays the evolution of the merger fraction of two separate subsamples for various magnitude and mass ranges. The overall sample, represented by the blue contours, shows relatively small variations in the merger fraction, making it challenging to discern a significant trend. Notably, for the Gold sample, the values remain stable across the entire range of redshifts under investigation. The calculated mean value, along with its associated error, is $f_m = 0.11\pm0.04$. This outcome aligns with previous research at lower redshifts \citep{Lotz_2006, Lin_2008, Conselice_2009a, Lòpez-Sanjuan_2009, Jogee_2009, Man_2016, Ventou_2017, Ventou_2019, Kim_2021}, all of which indicate a merger fraction around $f_m\sim0.10$ with lack of evidence for redshift evolution. We presented a simple argument based on dark matter halo merger rates and dynamical timescales of galaxies (which determine for how long a merger remains visible) to support a scenario where the merger fraction is almost redshift independent.\\
\indent $\bullet$ Our study aimed to investigate potential differences in the star formation rate between two categories: mergers and non-mergers. As illustrated in Fig.\ref{fig: sSFR vs z - hres - mu2.0}, we do not observe a clear and robust trend in the sSFR for mergers compared to non-merger galaxies, with redshift as the independent variable. Therefore, we conclude the sSFR is independent of whether a galaxy is a merger or not. In turn, this seems to suggest that smooth accretion (or minor mergers) are the main drivers in setting the SSFR for a star forming galaxy.\\
\indent $\bullet$ There is no apparent correlation between the identification of mergers according to the wavelengths of the observation filters utilized to establish the initial sample. Consequently, the merger fraction does not display any specific pattern in this aspect maintaining a level compatible with the main result of this work of $f_m=0.11\pm0.04$ across the examined redshift range. This result extends the findings obtained at lower redshifts, where \citealt{Treu_2023,Vulcani_2023} argue that morphological parameters, which form the basis for classifying mergers and non-mergers and subsequently calculating the merger fraction, do not exhibit a strong dependence on the redshift at which galaxies are observed.\\
\\
\indent To go beyond the preliminary results for the merger fraction during the initial stages of the epoch of reionization, further data is required. To improve statistical robustness and address potential cosmic variance effect \citep{TrentiStiavelli2008} as well as any weak-magnification lensing bias, a promising prospective strategy involves examining galaxies and images across diverse fields (e.g., COSMOS \citep{Scoville_2007}, EGS \citep{Davis_2007}, PRIMER \citep{Dunlop_2021}), in addition to those considered in this work. This will enable a thorough exploration of the galactic environment, incorporating factors such as metallicity and spatial density and open opportunities to quantitative comparison to predictions from numerical simulations and theoretical modelling of galaxy assembly in the first billion years.

\section*{Acknowledgements}

We thank the anonymous referee for useful suggestions and
comments that have improved the manuscript. This research was supported by the Australian Research Council Centre of Excellence for All Sky Astrophysics in 3 Dimensions (ASTRO 3D), through project number CE170100013. This work is based on observations made with the NASA/ESA/CSA James Webb Space Telescope. The data were obtained from the Mikulski Archive for Space Telescopes at the Space Telescope Science Institute, which is operated by the Association of Universities for Research in Astronomy, Inc., under NASA contract NAS 5-03127 for JWST. These observations are associated with programs JWST-ERS-1324, JWST-GO-2561, and JWST-DDT-2756. BV acknowledges support from the INAF Large Grant 2022 “Extragalactic Surveys with JWST” (PI Pentericci). BV is supported  by the European Union – NextGenerationEU RFF M4C2 1.1 PRIN 2022 project 2022ZSL4BL INSIGHT. MB acknowledges support from the ERC Advanced Grant FIRSTLIGHT and Slovenian national research agency ARRS through grants N1-0238 and P1-0188. BM acknowledges support from the Australian Government Research Training Program (RTP) Scholarship.\\

%%%%%%%%%%%%%%%%%%%%%%%%%%%%%%%%%%%%%%%%%%%%%%%%%%
\section*{Data Availability}
The data used to conduct the analysis are from: (i) GLASS
JWST-ERS-1324 \citet{Treu_GLASS_2022}, (ii) UNCOVER JWST-GO-2561
\citet{Bezanson_2022}, and (iii) the Director’s Discretionary Time
Program 2756 (PI Wenlei Chen). Photometric catalog provided by \citealt{Merlin_2022,Paris_2023}.



\bibliographystyle{mnras}
\bibliography{main} % if your bibtex file is called example.bib

%\begin{thebibliography}{99}
%\end{thebibliography}

%%%%%%%%%%%%%%%%%%%%%%%%%%%%%%%%%%%%%%%%%%%%%%%%%%

%%%%%%%%%%%%%%%%% APPENDICES %%%%%%%%%%%%%%%%%%%%%

\appendix

\section{Wavelength dependency study}\label{app: wavelength dependency study}

\begin{figure*}
\includegraphics[width=1.\linewidth]{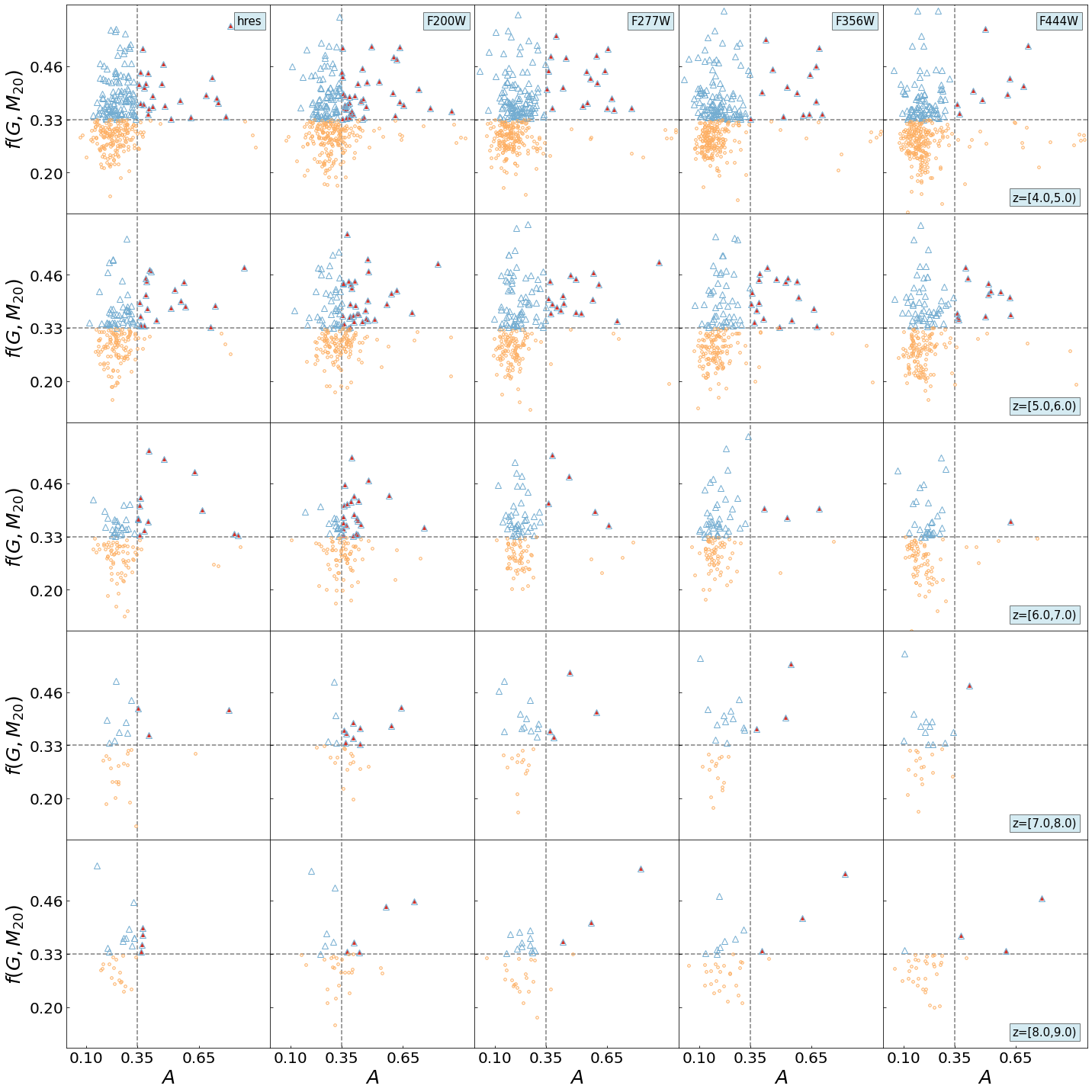}
\caption{Mergers classification method presented in Sec.\ref{subsec: merger identification}. Each row shows results for a different redshift bin, with each panel within each row showing results when a different filter from NIRCam is used for determining the morphological parameters. Galaxies located above the horizontal dashed line (representing Eq.\ref{eq: gini classical merger criteria}) are labelled with blue triangles and are classified in the Silver sample. Among these, the subset to the right of the vertical dashed line (representing Eq.\ref{eq: asimmetry classical merger criteria}) is highlighted in red, representing the galaxies in the Gold sample.}
\label{fig: GM20 vs A}
\end{figure*}

\begin{figure*}
\includegraphics[width=0.8\linewidth]{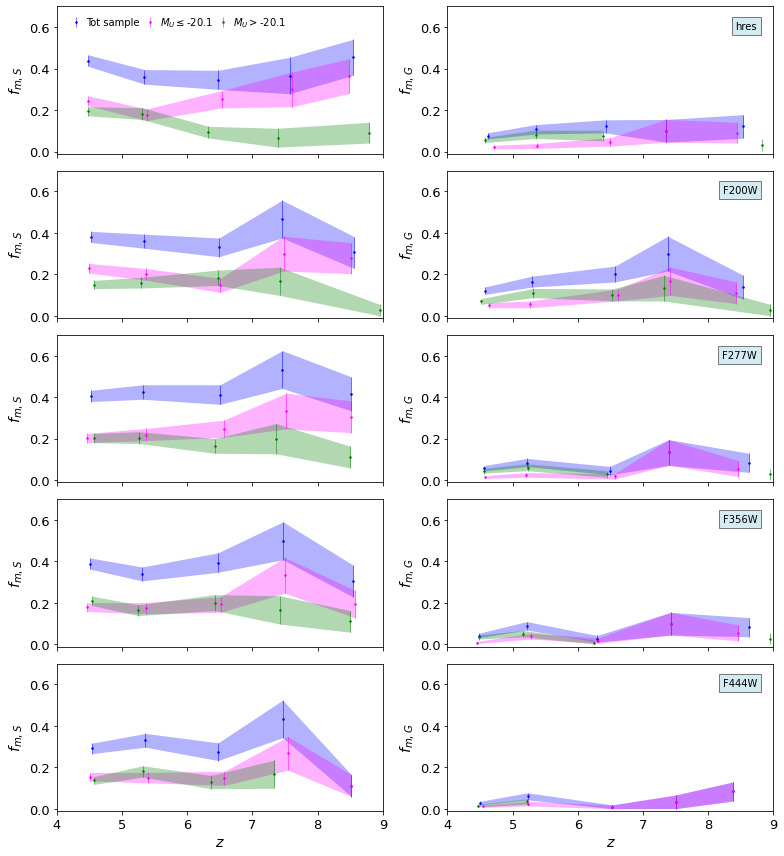}
\caption{Merger fraction evolution in respect to redshift, using different filters on the NIRCam instrument for the identification. The left panels show the Silver sample and the right panels represent the Gold sample. In each panel, we also show the results for the subsamples of bright and faint galaxies, defined by a cut in magnitude at $M_{U} = -20.1$. }
\label{fig: fm vs wave mag z fixed}
\end{figure*}

\begin{figure*}
\includegraphics[width=0.8\linewidth]{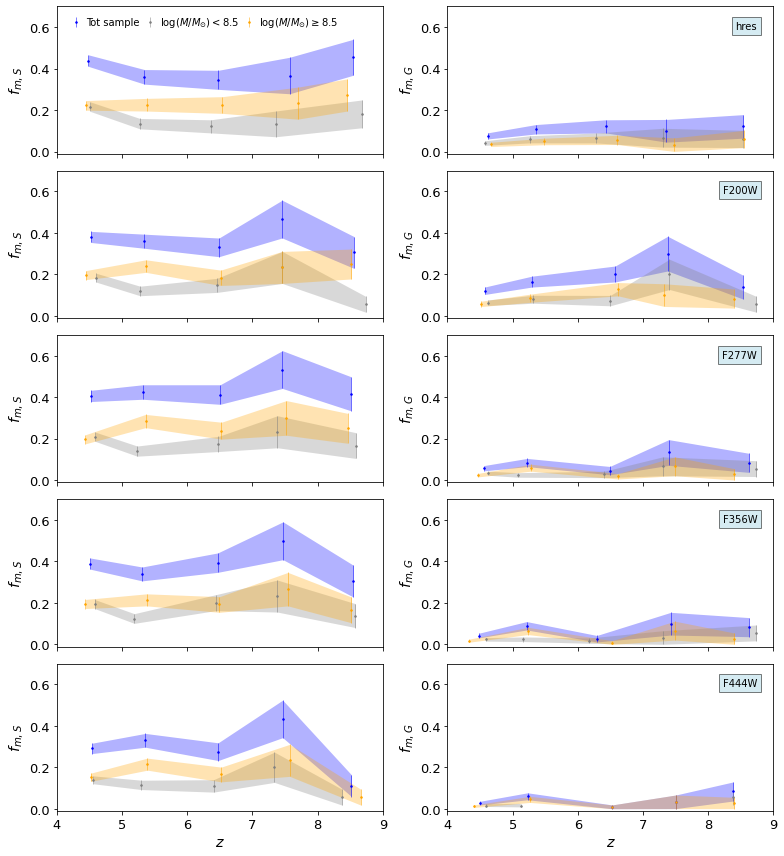}
\caption{As in Fig.\ref{fig: fm vs wave mag z fixed}, but splitting the parent sample into two subsamples based on stellar mass, with high-mass galaxies defined to be those with stellar masses greater than $M_{*} = 10^{8.5}M_\odot$, and low-mass galaxies having stellar masses below this value.}
\label{fig: fm vs wave mass z fixed}
\end{figure*}

In this Appendix, we investigate potential correlations between the detection of galaxy mergers and variations in the merger fraction $f_m$. This examination takes into account the various wavelengths used for candidate identification, utilizing a selection of bands, specifically F200W, F277W, F356W, and F444W, in conjunction with the \texttt{hres} band used as the reference in the main study.\\
\indent Fig.\ref{fig: GM20 vs A} provides a visualization of the morphological parameters within the same redshift bin and band scheme. Furthermore, Fig.\ref{fig: number G, GS, tot} offers a graphical representation of the number of galaxies within the total, Silver, and Gold samples. From these figures, it becomes apparent that there is no significant discrepancy in the number of galaxies detected in each band when examining the same redshift bin. This suggests that there is no pronounced trend in detection based on wavelength.\\
\indent To further explore potential dependencies, we applied the same magnitude and stellar mass cuts that were considered in the primary study. The goal was to assess whether the wavelength used for the survey impacted the results. Fig.\ref{fig: fm vs wave mag z fixed} and Fig.\ref{fig: fm vs wave mass z fixed} present the outcomes of this supplementary analysis. As demonstrated in these figures, the merger fraction $f_m$ does not show significant variations with the slope of the evolution trend with respect to redshift constrained within the range (0,0.02) for each band. The mean values of the merger fraction derived from the five different bands are: $f_m$[\texttt{hres}]$=0.11\pm0.04$, $f_m$[F200W]$=0.18\pm0.04$, $f_m$[F277W]$=0.09\pm0.03$, $f_m$[F356W]$=0.07\pm0.04$ and $f_m$[F444W]$=0.5\pm0.03$. Overall, most of the results agree within $1\sigma$ with the primary result obtained using the \texttt{hres} band of $f_m=0.11\pm0.04$, while the remaining results fall within the $1.5-2\sigma$ range.\\
\indent In summary, our analysis reveals that there is no strong increasing or decreasing trend in the merger fraction when considering different detection wavelengths. These results are consistent with the previous findings by \citealt{Treu_2023,Vulcani_2023}, which suggested that the morphology of Lyman Break Galaxies remains relatively consistent across different wavelengths, from the rest frame optical to the rest frame UV. 

\begin{figure*}
\includegraphics[width=1.\linewidth]{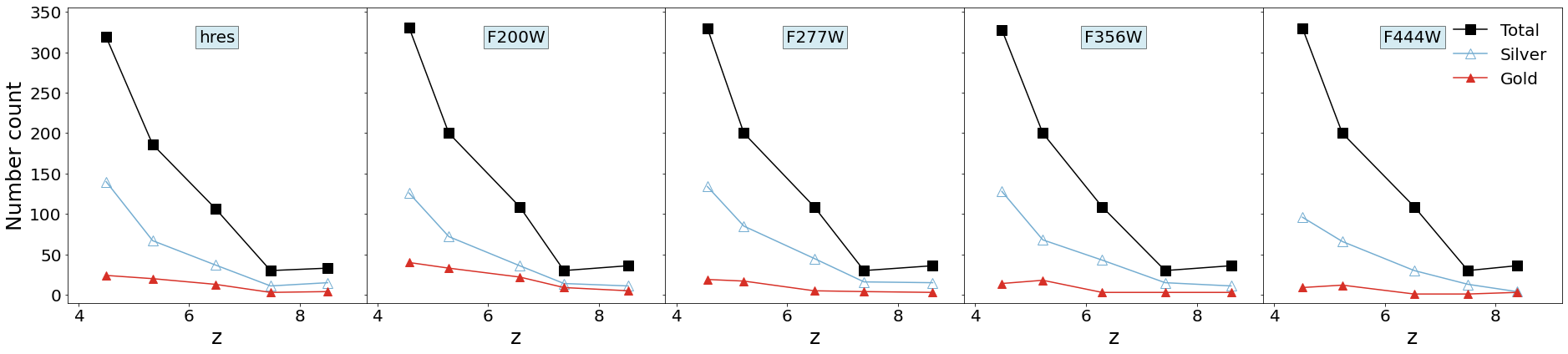}
\caption{Demography for the galaxy (merger) samples versus redshift. Points correspond to those in Fig.\ref{fig: GM20 vs A}. Black squares indicate the total number of candidates considered in each redshift bin. Blue triangles represent galaxies in the Silver sample, while red triangles denote those classified in the Gold sample.}
\label{fig: number G, GS, tot}
\end{figure*}

%%%%%%%%%%%%%%%%%%%%%%%%%%%%%%%%%%%%%%%%%%%%%%%%%%

% Don't change these lines
\bsp	% typesetting comment
\label{lastpage}
\end{document}